\documentclass[12pt]{article}
\begin{document}
\baselineskip24pt
\begin{center}
{\large\bf GRAVITY SANS SINGULARITIES} 

\medskip 

K. H. Mariwalla\footnote{{\em E-mail}:~mari@imsc.ernet.in} \\
{\em Institute of Mathematical Sciences \\
C.I.T. Campus, Tharamani, Chennai - 600113, INDIA} 
\end{center}

\centerline{\bf Summary} 

\smallskip 

\noindent 
Basis and limitations of singularity theorems for Gravity are examined.  As 
singularity is a critical situation in course of time, study of time paths, in 
full generality of Equivalence principle, provides two mechanisms to prevent 
singularity.  Resolution of singular Time translation generators into space of 
its orbits, and essential higher dimensions for Relativistic particle 
interactions has facets to resolve any real singularity problem.  Conceptually, 
these varied viewpoints have a common denominator: arbitrariness in the 
definition of `energy' intrinsic to the space of operation in each case, so as to 
render absence of singularity a tautology for self-consistency of the systems. 

\newpage


\noindent {\bf 1}. Following powerful theorems in General Relativity (GR) 
gravitational singularities have acquired rare acceptability, unusual in 
physics.  Singularity in a theoretical framework implies a critical situation in 
course of time.  Relativity introduces complications of light cone (null paths, 
surfaces) and inseparability of Space from Time (causality, Hyperbolicity).  
Taking gravity as a central force, a null surface $\vec{X}^2 - t^2$ $=$ $0$ 
defines for a spatial section the metric $ds^2$ $=$ $d\vec{X}^2 - dt^2$ $=$ 
$R^2d\Omega_2^{\ 2}$; so, as a body contracts under gravity beyond $r < R$, a 
spherical 2-surface gets `trapped'.  This forms basis of singularity theorems, 
beside sufficiency conditions for a minimum for proper time path functional
$\int d\tau$, throughout.  For not too specialised (relative to curvature 
components) velocity field $v$, this is better studied with the (Hamilton-Jacobi 
analogue) Raychaudhuri equation: $\tau$-rate trace of $v$-Lie derivative of the 
metric ($2\dot{\theta}$) is quadratic in velocities with Ricci tensor 
coefficients (plus terms in $-(\nabla v)^2$ and $\nabla\dot{v}$); via Einstein 
equation, one obtains for `positive energy condition', inevitability of 
singularity \cite{H}.  Despite, generality of this and related theorems, there 
are inherent limitations, as path or velocity field refers to an ideal 
test-particle: structureless, inviting no reaction, {\em e.g.}, in a closed system 
in mechanics, centre of mass may be fixed, but angular position can vary {\it ad 
libitum}; spin plays a crucial role here, and is ignored.  Also ignored are 
forces (like Coriolis-magnetic field) that do no work - the only nongravity 
relativistic forces, giving added dimension to `interaction space', where 
interaction energy `resides', just as gravitational energy resides in 
space-time.  Resolution of incomplete and non-global Hamiltonians in mechanics 
are also relevant.  \\   

\noindent {\bf 2}.  Einstein \cite{E} envisaged geometric Gravity as a 
generalization of Special Relativity (SR), in that: by weak principle of 
equivalence a test particle follows a geodesic, and by strong principle the 
metric is a covariant constant; both allow torsion.  In geometric terms, the 
first law in mechanics expresses vanishing of curvature and torsion; while a 
spherical rotator corresponds to vanishing curvature (not torsion) only; this 
yields \cite{M1} its symmetry group as $SO(n+1)$ and spin-torsion relation.  

A geodesic equation admits a further symmetry 
$\Gamma^\mu_{\nu\sigma}$ $\longrightarrow$ 
$\Gamma^\mu_{\nu\sigma}$ $+$ $\delta^\mu_\nu \lambda_\sigma$; 
its antisymmetric part contributes to torsion to define precise spin-torsion 
relation in Einstein-Cartan theory \cite{T}.  The symmetric part redefines the 
path parameter. An integrable change $\lambda_\sigma$ $=$ $\partial_\sigma 
\lambda$, gives Einstein's \cite{E} $\lambda$-transformation leaving curvature 
(with torsion) unchanged; Weyl's projective curvature is also unchanged, implying 
a same system of paths.  Projective Einstein-Cartan structure is the most 
general, consistent with Equivalence principle and Einstein equations.  The 
energy tensor gives only the contribution of matter, which balances or complements 
the gravitational field, as expressed in space-time geometry.  Due to internal 
changes - no external energy is supplied - two sides of Einstein equations may, 
in principle, undergo spontaneous change governed by spin-torsion relation
in Einstein-Cartan theory, or by projective change relating two (symmetric) 
connections to provide for singularity avoidance: 
\begin{itemize}
\item[({\it a})] 
In a closed gravitating system a continual process of capture and expulsion is on,
involving interconversion of orbital and spin angular momenta, and spin
redistribution; it may be looked upon to generate torsion, leaving paths 
unchanged, but redefine Einstein and Energy tensors.  In Raychaudhuri equation 
such a change is known \cite{T} to provide for avoidance of singularity. 
\item[({\it b})] 
A spontaneous projective change, alters the (covariant) Ricci tensor additively 
\cite{M2} and may be used in Raychaudhuri equation to avoid singularity.  
Elsewhere we have treated Robertson-Walker metric for imploding cold body for 
singularity avoidance \cite{M3} and for Hawking radiation \cite{M2}.  But the 
present argument is more general.  
\end{itemize} 

\medskip 

\noindent {\bf 3}. In mechanics for linear force the orbits are either only 
closed or open; the corresponding Schr\"{o}dinger operators have respectively 
discrete or continuous spectra; here Hamiltonians generate finite canonical 
transformations and belong to $sl(2;R)$ Lie algebra.  For the Kepler problem, 
Hamiltonian vector field is known to be {\em incomplete} and non-global.  By a 
change in time parameter, it is resolvable into $1$-parameter subgroups of 
$sl(2;R)$, that have either discrete or continuous spectra.  For Hamiltonians 
with more complicated spectra of uneven spread of point and continuous
spectrum, one expects some representation of Virasoro algebra.  For path-integral
computations and uncertainty relations, only such globally Hamiltonian vector 
fields will do. {\em Thus incompleteness does not necessarily imply a 
singularity, when it can be resolved}.  Direct application of this to GR 
singularities is perhaps problematic, as space and time are well entwined.  But, 
there is a connection between approaches based on curves and $1$-parameter groups 
via discrete subgroups of conformal group, implicit in Poincare's work on 
automorphic functions and bounded domains.  \\ 

\noindent {\bf 4}.  Relativistic Kinematics of particles describes all 
elastic/inelastic processes without a potential, as all interaction energy comes 
from the masses.  For translation generators $P_\mu$, $mv_\mu$ $\longrightarrow$ 
$P_\mu - A_\mu$ defines the displacement field $A_\mu$ as carrier of interaction 
energy modes, corresponding to forces (like Coriolis) which do no work.  Taken as 
matrix valued - say, in basis of adjoint representation of a Lie algebra $g$ of a 
compact group ${\cal G}$ - it defines a vector space $V^n$ added on to the space 
of tangents at each point of $M^d$, to give interaction space $M^{d+n}$ $=$ $M^d 
\times {\cal G}$ locally.  Since, for $m \geq 0$, the fields are defined over 
doublesheeted hyperboloid/cone, a particle may be conceived of as surrounded by 
fluctuating clouds of virtual particles and antiparticles, as well $A_\mu$ modes 
in which they may unite to go back and forth.   {\em We now make a drastic 
assumption}: the entire energy $m$ of a typical particle is determined by energy 
of a `mean' Newtonian oscillator relative to proper time $\tau$; {\it viz}.  
$$ 
\frac{1}{2}\mu\left(\frac{d\xi}{d\tau}\right)^2 + \frac{1}{2}\omega^2\xi^2 
 = \frac{1}{m}P_\mu P_\mu = m = m \frac{dt^2-d\vec{X}^2}{d\tau^2}\,.  
$$
For time-fluctuation $\xi$ $=$ $t$, de Sitter space obtains as relativistic 
analogue of uniform acceleration.  For spatial-fluctuation in $N$-dimensions, one 
obtains $5$-dimensional anti-de Sitter (Ads) space $\times$ $S^{m-1}$.  Spatial 
part of Ads, in these coordinates, is a space of negative constant curvature, 
rewritten as $$ds_4^2 = R^2\left[ d\alpha^2 + (\sinh\alpha)^2 d\Omega_3^{\ 2} 
\right]. $$ 
Analogue of Coulomb potential in such $N$-space is 
$$
\psi_N = \int({\rm cosech}\,\alpha)^{N-1} d\alpha \ \ 
\stackrel{r = \alpha R \ \ R \rightarrow \infty}{\longrightarrow} \ \  
-\frac{mGR^{N-3}}{r^{N-2}}. 
$$ 
For $N = 3$ it is $(1-{\rm coth}\,\alpha)$ $\longrightarrow$ $-MG/r$; the 
Schr\"{o}dinger operator in space of negative curvature with this potential has 
finite number of eigenvalues; and as model of `black hole interior' lowest 
eigenvalue gives measure of (BH entropy)$^2$ \cite{M4}.  For $N = 4$ for large 
$R$, potential is $\lambda r^{-2}$, for which the flat space Schr\"{o}dinger 
operator has $s$-state eigenvalues \cite{C} $$-E_n \sim \exp[2B - 
(2n+1)\pi(4\lambda-1)], $$ where each real $B \neq 0$ defines a new self-adjoint 
extension of Hamiltonian, with $B$ $\longrightarrow$ $B+M\pi$ $(n,M = \pm\,{\rm 
integers})$ leaving all unchanged.  A particular disadvantage is that point 
spectrum extends from $-\infty$ to $0$.  Consider now the bundle of all such 
extensions, and pick a subset where $M$ and $2n+1$ adjust to give a finite lower 
bound (even degenerate) with proper orthogonality (and completeness, with 
continuous spectrum) preserved.  {\em Thus particles sit here confined, as if 
free, but in a higher dimensional space with strong gravity}; and singularity is 
avoided.  For the case of the phase operator, bundle of all its extensions give 
\cite{M5} the $kq$-representation of Zak.  Such interpretation is yet unavailable 
for $B$.  \\

All four mechanisms cited (spontaneous spin/torsion or projective change; non-global,
incomplete Hamiltonians; Relativistic interaction space), enabling avoidance of
singularity, have a common denominator - arbitrariness in quantifying available
interaction energy ({\em viz}. gravitational in GR, mechanics, and particle 
interactions) inherent in the definition of space in which the energy `resides'.  
In that sense it is a tautology and evidence of consistency of the systems 
considered. \\ 

\noindent {\em Acknowledgements}: This work forms part of a larger study of 
Origin and Structure of Time to appear in the e-print Archive.  The ideas in Sec. 
4 were originally floated at Boulder in 1970.  Thanks are due to R. Jagannathan 
for invaluable help.


\bigskip 

\begin{thebibliography}{99}
\bibitem{H} 
S. W. Hawking, G. F. R. Ellis, {\sl The Large Scale Structure of Space-Time} 
(Cambridge Univ. Press, 1973). 
\bibitem{E}
A. Einstein, {\sl Meaning of Relativity}, 6th Ed. (Mathuen, 1956). 
\bibitem{M1}
Kishin Mariwalla, {\em Phys. Rep.} {\bf 20C} (1975) 287. 
\bibitem{T}
A. Trautman, {\em Nature} {\bf 242} (1973) 7.   
\bibitem{M2}
K. H. Mariwalla, {\em Phys. Lett.} {\bf 68A} (1978) 409. 
\bibitem{M3} 
K. H. Mariwalla, ``Geometrization of Creation Field'', in {\em Proc. Conf. 
Cosmology and Gravitation}, MATSCIENCE Report {\bf 84} (1971) 83. 
\bibitem{M4} K. H. Mariwalla, ``A Model of Interior of a Black Hole'', in 
{\em Proc. Tamilnadu Acad. Sci.} {\bf 5} (1982) 83. 
\bibitem{C}
K. M. Case, {\em Phys. Rev.} {\bf 80} (1950) 797. 
\bibitem{M5} 
K. H. Mariwalla, ``Complementarity and Coherence'', in {\em Proc. Tamilnadu 
Acad. Sci.} {\bf 2} (1979) 107.  
\end{thebibliography}
\end{document}